\documentclass [10pt]{article}

\usepackage[makeroom]{cancel}
\usepackage{mathpazo}

\usepackage{amsmath}

\usepackage{slashed}

\usepackage{hyperref}
\usepackage{geometry}

\usepackage{graphicx}
\usepackage{amssymb}
\usepackage{epstopdf}

\usepackage{cite}
\usepackage{graphicx}
\usepackage[usenames]{color}
\usepackage{subfigure}
\usepackage{setspace}
\usepackage{slashed}
\usepackage{multirow}
\usepackage{ulem}

\usepackage[hang,small]{caption}

\usepackage{geometry}
    \usepackage{amsmath}   
        \usepackage{amssymb}   
    
    \newcommand{\ignore}[1]{}
 \newcommand{\ee}{\end{equation}}

\newcommand{\ab}{\allowbreak}

\def\ba#1\ea{\begin{align}#1\end{align}}
\newcommand{\bit}{\begin{itemize}}
\newcommand{\eit}{\end{itemize}}
\newcommand{\im}{\item}

\newcommand{\nn}{\nonumber} \renewcommand{\bf}{\textbf}
\newcommand{\ra}{\rightarrow}

\newcommand{\p}{\partial}

\usepackage{fancyhdr}
\newcommand{\iu}{{i\mkern1mu}}
\usepackage{titlesec,titletoc}
  \titleformat{\section}{\Large\sf\bfseries}{\thesection}{1em}{}
  \titleformat{\subsection}{\large\sf\bfseries}{\thesubsection}{1em}{}

\title{\sf\bfseries \ntitle}
\author{
    Sumeet Dagaonkar\footnote{sumeetkd@iitk.ac.in}~, Pankaj Jain\footnote{pkjain@iitk.ac.in}~\\
    \it{Department of Physics, Indian Institute of Technology, Kanpur 208016, India}\\
\and 
John P. Ralston\footnote{ralston@ku.edu}\\
\it{Department of Physics \& Astronomy, University of Kansas,}\\
\it{Lawrence, KS - 66045, USA}\\
}

\date{}%{\today}

 \newcommand{\pghdr}{\footnotesize {S. Dagaonkar} {\it et al.} -- Uncovering the Scaling Laws \dots }
\newcommand{\ntitle}{Uncovering the Scaling Laws of Hard Exclusive Hadronic Processes in a Comprehensive Endpoint Model}

\begin{document}

\vspace{-3cm}

\maketitle
\begin{abstract}{We show that an endpoint overlap model can explain the scaling laws 
observed in exclusive hadronic reactions at large momentum
transfer. The model assumes  one of the valence quarks carries most of the hadron momentum. Hadron form factors and fixed angle scattering are related directly to the quark wave function, which
can be directly extracted from experimental data. A universal linear endpoint behavior explains the proton electromagnetic form factor, proton-proton fixed angle scattering, and the $t$-dependence of proton-proton scattering at large $s>>t$. Endpoint constituent counting rules relate the number of quarks in a hadron to the power-law behavior. All proton reactions surveyed are consistent with three quarks participating. The model 
is applicable at laboratory energies and does not need assumptions of asymptotically-high energy regime. 
A rich
phenomenology of  lepton-hadron scattering and hadron-hadron
scattering processes is found in remarkably simple relationships between diverse processes.
  }
\end{abstract}
\vspace{-0.6cm}

\section{Experimental Regularities}

The experimental study of differential cross sections of hard exclusive hadronic reactions
at high energy reveals a remarkable pattern: {\it They are described by power laws} \cite{GRF,matveev,Sivers76}. A model
explanation exists \cite{Farrar79, BL80,Radyushkin80,Radyushkin80a}, yet it is not satisfactory \cite{Isgur} at the energies of experimental measurements. We are driven to find a consistent explanation of experimental regularities by re-examining all the facts from a fresh point of view.

``Hard'' reactions are those which depend on a single large scale $Q^{2}> GeV^{2}$, or several large scales with a fixed ratio. It is remarkable that the proton electromagnetic form factor $F_{1}(Q^{2})$ agrees well with a decreasing power of $Q^{2}$ for $Q^{2} \gtrsim 5 \, GeV^{2}$ \cite{Andivahis}. For large momentum transfer, it is remarkable that $pp \ra pp$ fixed-angle cross section $d \sigma/ dt$ agrees well with a decreasing power of $Q^{2}\sim s$ \cite{Landshoff_Pol}, where $\sqrt{s}$ is the center of mass energy. There are many other examples. 

We have re-evaluated the phenomenology of power-law dependence or 
``scaling laws'' for exclusive reactions. Due to history, the most simple and plausible explanation failed to be developed. The model appears in the literature as the Feynman process, also known as the Drell-Yan model, also known as {\it the endpoint overlap model} 
\cite{Feynman69,Drell70,West70}. There is
actually much to be learned and much that is new when the model is objectively explored. 

\subsection{The Endpoint Overlap Model }

In their evaluation of the endpoint region, Brodsky and Mueller\cite{Mueller}
wrote that ``its contribution depends sensitively on the hadronic wave functions''. The discussion discovered no actual fault in the endpoint contribution. Instead of finding a flaw, the section ends with a weak suggestion to assume validity of a short-distance perturbative model, as ``at least plausible'', adding {\it ``in any case there is currently no comprehensive alternative theory of these processes''.} 

We suggest that the lack of a comprehensive alternative theory came out of {\it a historical failure of the endpoint contribution to be fully appreciated and developed.} In the concluding remarks we link this to the history of early development of perturbative QCD, which is an era long past. 

In reviewing the current status we noticed several facts: 
\bit \im The predictions of all models depend on the wave functions. For reasons we believe are obsolete, the opportunity to learn about wave functions using data was bypassed in the promotion of short-distance ($SD$) models \cite{Farrar79, BL80,Radyushkin80,Radyushkin80a}. 
\im Great emphasis was imposed early on asymptotic limits. The motivation was not to learn about hadrons, but an attempt to make hadrons irrelevant for the goal of establishing QCD. 
\im The asymptotic limits of QCD are now understood to be of
negligible experimental relevance. The asymptotic limits of QCD predictions have also never actually been established.  Instead limits of models have been established. None of the work assuming a model has gone an inch beyond the boundaries of the model itself. In particular, it has not been shown anywhere that the pion form factor {\it of the general theory known as QCD} necessarily falls faster than $1/Q^{2}$, despite considerable effort to force such a conclusion. Careful reading is needed to verify this. For example, 
Farrar and Jackson \cite{Farrar79} claim an asymptotic limit in opening lines, without actually supporting the claim: The contrary information about regions outside the model is buried in the footnote labeled Ref. 13 in the paper.\im 
Now that QCD is established, every integration regime, including those contradicting the assumptions of $SD$ models, needs to be considered. Interest in the larger theory and hadron structure has eclipsed the goal of exhibiting 
a model based solely on perturbation theory. \im The main reason for early interest in perturbative models was power-law behavior. Inexperience with more general models created a folk-lore that ``soft'' non-perturbative wave functions would lead to {\it exponential } dependence on a large $Q^{2}$ scale. This is false. As we review, power-law behavior is generic from the endpoint region.  \im Divisions in the field separated groups into two camps. Relying on perturbation theory appears to be more theoretically ambitious, but it is actually less general than representing dynamics with wave functions. For one thing, an arbitrary order of perturbation theory can be subsumed into equivalent wave functions, but not vice versa.\im Calculations in perturbation theory use the Fock state basis of free field theory. Considerable effort has been dedicated to making the endpoint region of perturbative calculations go away at asymptotically high energies, towards demonstrating {\it perturbative} self-consistency. None of that work is relevant to non-perturbative wave functions, which use a different basis of fully-interacting quanta. It is not logically self-consistent to extend the asymptotic pQCD-based suppression of the endpoint region to wave functions extracted from experimental data. \im Despite years of study, very little is known about pions and protons. The proof comes from the dearth of definite information about pions and protons in terms of non-perturbative wave functions. Contrary to the bias of perturbative QCD, it is {\it definitely possible and absolutely productive} to use experiments to learn about non-perturbative wave functions.\eit

Recently Chang, Clo\`et, Roberts, Schmidt, and Tandy 
\cite{CCRST1,CCRST} have computed the pion form factor with a method described as self-consistent for all space-like $Q^{2}$. The paper highlights the asymptotic $SD$ model's prediction being about 3 times smaller than the experimental form factor. Ref. \cite{CCRST} states the asymptotic estimate is incapable of converging to a realistic value below $Q^{2} \sim 1000 \, GeV^{2}$. This is typical of asymptotic $SD$ estimates: The estimates require fabulously high momentum transfer to apply. We agree that the assumptions made in setting up $SD$ models are contradicted by the application of the models to existing momentum transfers. 

The best test of the $SD$ models comes from hadronic helicity conservation\cite{helicityconservation}. These tests are much less demanding than asymptotic limits, and apply to an expansion of leading power behavior. The tests fail in almost every case experiments exist \cite{Preparata79,Brodsky79,GP95,OFallon77,Lin78,Linn82,Aschman77,Abe76,Antille81,Hansen83,Crabb78,Crosbie81,Peaslee83,Khiari89,Wijesoriya,Gilman02,Jones00}. That is convincing evidence that the observed experimental regularities are not explained by the $SD$ model.

The calculations of \cite{CCRST} are made in an overlap model emphasizing first-principles predictions of the pion wave function. They are similar in spirit to overlap models in the relativistic impulse approximation \cite{Gross14a,Gross14b,Gross12}. These models 
are very successful in describing data at low $Q^{2}$. Our approach accepts the validity of these models as integral representations of form factors, while also extending the scope to other processes. We differ from most models by not attempting to know wave functions in advance. Many studies have been restricted to estimating the endpoint region to order of magnitude. We will show that the end-point region of the wave functions
is not only determined, but over-determined, by
experimental regularities found in power laws.
By {\it measuring} the wave functions rather than predicting them we find the endpoint overlap model is a consistent comprehensive description. Rather unexpectedly, our approach extracting information from wave functions is quite consistent with the trend predicted by \cite{CCRST}. 

Section \ref{sec:counting} derives the {\it endpoint overlap constituent counting rules} for form factors. Section \ref{sec:hadrons} extends the rules to exclusive hadron-hadron reactions. These rules explain why scaling laws should be observed at the limited energies of laboratory experiments, and how scaling laws are correlated with the number of quarks scattering. We cannot explain why this predictive regularity of {\it all exclusive reactions surveyed} has been overlooked. Concluding remarks with brief historical commentary are given in Section \ref{sec:conclude}.

\section{Endpoint Power Counting Rules}
\label{sec:counting}

In this Section we derive the {\it endpoint constituent counting rules}, which predict the scaling power of $Q^{2}$ in terms of the number of constituents. Comparing experimental data to these rules finds consistency with three (3) quarks scattered in every proton reaction we have surveyed. The discussion will be organized in increasing levels of detail, beginning with the simplest case of the pion electromagnetic form factor. 

\subsection{The Pion Form Factor $F_{\pi}(Q^{2})$: The Probability of a Slow Quark} \label{sec:fpi}

\begin{figure}[htbn]
\begin{center}
\includegraphics[width=4in]{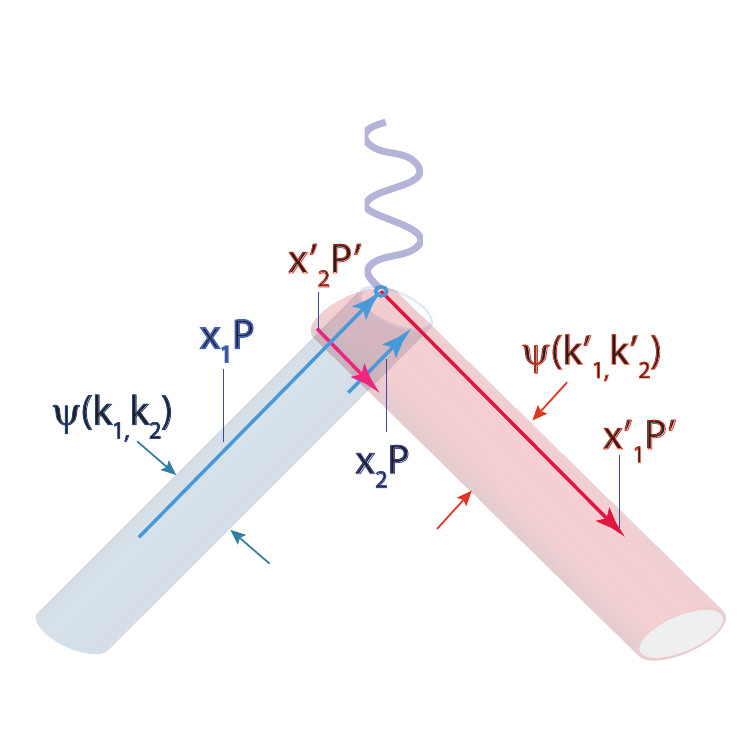}
\caption{ \small Physical picture of the endpoint dominance model.  Due to the change of direction of the fast momentum, the transverse momentum scale of non-perturbative wave functions must overlap with the range of $x_{2}P/x'_{2}P'$ of spectator constituents. The coordinate system of variables $k_{i}$ and $k'_{i}$ are defined in the text.  }
\label{fig:EndpointPic}
\end{center}
\end{figure}

The form factor is defined by \ba  <P'|J_{em}^{\mu}|P> = (P+ P')^{\mu } F_{\pi}(Q^{2}), \ea where $J_{em}^{\mu}$ is the electromagnetic current operator. In a gauge-invariant local field theory the photon interaction involves one (1) struck parton. The minimum number of constituents in the pion is two. The dynamical question of elastic scattering is how scattering one constituent can scatter the entire hadron. This is answered by the quantum mechanical overlap of wave functions.
Figure \ref{fig:EndpointPic} conveys the qualitative picture. The endpoint region is dominated by some transverse hadronic scale ``$\Lambda$'' for which the slow parton obeys $x_{2} \lesssim \Lambda/Q$, while the struck parton obeys $1-x_{1}\lesssim\Lambda/Q$. For the entire region the endpoint contribution is such that the transverse momentum integrations do not contribute any power of $Q^2$ to the form factor. This happens to be the feature causing the short-distance model to be impossible to justify in this region. 

The form factor is 
\ba F_\pi(Q^2) & = \int [dk ] [dk'] \,  \psi^{'*}(x'_{i}, \, \vec{k}'_{i}  )  T \psi(x_{i}, \, \vec{k}_{i}),  \nn \\
& [dk]=dx_{1}dx_{2}dk_{-1}dk_{-2}d^{2}k_{1}d^{2}k_{2} \nn  \ea  The electromagnetic interaction matrix element $T$ contains the quark charges, a gamma matrix, and momentum-conserving delta functions. Wave functions with several independent spin structures are possible, with discussion postponed to Section \ref{sec:gauge}. We define the delta functions to include factors representing momentum conservation, hence $\vec k_{1}+\vec k_{2}=\vec P$, etc. in a frame we now specify.

Let $P, \, P'$ be the 4-momenta of the pions, with $P'=P+q$, $q^{2} =-Q^{2}$. Choose a Lorentz frame with Cartesian labels $(E, \, p_{ X}, \, p_{Y}, \, p_{Z})$ where $E$ is the energy. Thus \begin{equation}\begin{split}  q &= (0, \, Q, \, 0, \,0) ;   \\ 
P &= (\sqrt{Q^{2}/2+m_{\pi}^{2}}, \, -Q/2, \, 0, \,Q/2 ) ;   \\ P' &=(\sqrt{Q^{2}/2+m_{\pi}^{2}}, \, Q/2, \, 0, \,Q/2 ).  \label{eq:frame}\end{split}\end{equation}

\begin{figure}[htbp]
\begin{center}
\includegraphics[width=3in]{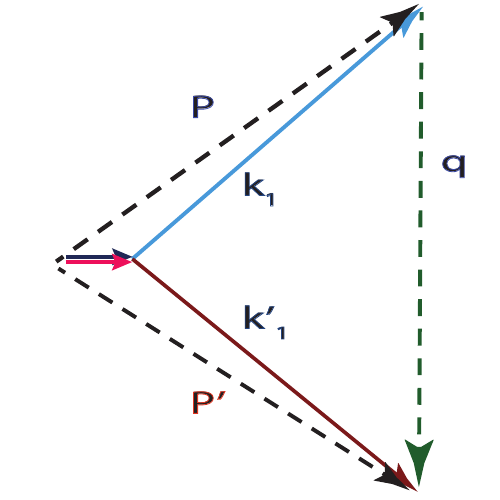}
\caption{ \small Endpoint kinematics in the pion case. Pion momenta are shown as dashed arrows, while quark momenta are solid arrows. Two isosceles triangles representing energy and momentum conservation must close. The transverse and longitudinal momenta of one spectator covers the difference. By inspection, $ k_{x}=xQ/2 $, where $x$ is the momentum fraction of the slow quark.  }
\label{fig:EndPion.pdf}
\end{center}
\end{figure}

Let $k^{\mu}_{1}$  ($k^{'\mu}_{1}$) be the momenta of the struck parton before (after) scattering. Due to the change of direction of the fast momenta, the meaning of symbols ``$\vec{k}_{i}$'' must be adapted to be orthogonal to each hadron's direction. We introduce a basis of transverse vectors adapted to the particular hadronic momenta : \ba \hat y =&(0, \, 0, \, 1, \,0) =\hat y'; \quad \vec P\cdot \hat y=\vec P' \hat y'=0; \nn \\ \hat x =& (0, \, -1, \, 0, \, -1), \quad \vec P\cdot \hat x= 0;\nn \\ \hat x' =& (0, \, 1, \, 0, \,-1),
\quad \vec P'\cdot \hat x'=0. \nn \ea With these coordinates, the components of the quark momenta are \ba k_{i}& = x_{i}P+ k_{ix}\hat x +k_{iy}\hat y =(x_{i}\sqrt{Q^{2}/2+m_{\pi}^{2}}, \, -x_{i}Q/2, \, 0, \, x_{i}Q/2) + (0, \, -k_{ix}, \, k_{iy}, \, -k_{ix}); \nn \\ 
k'_{i}& =x'_{i}P'+ k'_{ix} \hat x'+ k'_{iy} \hat y= (x_{i}\sqrt{Q^{2}/2+m_{\pi}^{2}}, \, x_{i}Q/2, \, 0, \, x_{i}Q/2  ) + (0, \, k'_{ix}, \, k'_{iy}, \, -k'_{ix}).\label{eq:qmomenta}\ea There are only three free parameters, and the quanta are {\it not} strictly constrained to the perturbative mass shell. By hypothesis, amplitudes are concentrated in the kinematic region shown. Integrating over a fourth (minus, or virtuality parameter) concentrated on the region is equivalent. Momentum conservation of the un-struck spectator is $k_{2}=k_{2}'$. Momentum conservation of the struck quark $k'_{1} =k_{1}+q$ yields four constraints: \ba & \quad k_{1x} = k'_{1x} \equiv k_{x}; \quad k_{1y} = k'_{1y} \equiv k_{y}; \nn \\  
     & x_{1}=x'_{1}; \quad x'_{1}Q/2 +k'_{1x} = -x_{1}Q/2+Q-k_{1x} \label{sum} \ea  Solving gives \ba  k_{x}  =  \frac{(1-x_{1})Q} {2} =\frac{x_{2} Q} {2}  . \nn \ea See Figure \ref{fig:EndPion.pdf}. We emphasize these are exact kinematic relations of the model, regardless of the value of $m_{\pi}$.

Evaluating $F_{\pi}$ gives \ba F_\pi(Q^2) =& \int_{0}^{1} dx    \, \Phi_{\pi}(x, \, xQ/2); \label{phihere} \\  \Phi_{\pi}(x, \, k_{x}) &=  \int dk_{y}\psi^{'*}(x , \, k_{x} , \, k_{y}  ) \psi(x , \,k_{x} , \, k_{y}). \label{basic} \ea  
A soft non-perturbative wave function means that $ \Phi_{\pi}(x, \, k_{x})$ is a rapidly falling function with a scale $k_{x} \lesssim \Lambda$, where $\Lambda \sim 300 $ MeV. To see how power-law behavior emerges, consider an exponential function multiplied by a function of $x$: 
\begin{equation}
  \Phi_{\pi}(x, \, k_{x}) = e^{-|k_{x}|/\Lambda}\phi(x); \label{ans} \end{equation} 
\begin{equation}
F_\pi(Q^2) = \int_{0}^{1} dx     \,  e^{-xQ/2\Lambda}\phi(x).  \end{equation} 
As $Q \ra \infty$ the exponential is driven toward the
endpoint $x_{2} \lesssim \Lambda/Q$.

Integrals dominated by their endpoints\footnote{The same expansion applies to a wide class of integrals with saddle points in $x$ approaching the endpoint} have an asymptotic series expansion\cite{bender} developed with integration by parts. The first term is given by 
\begin{eqnarray}  F_{\pi} &=& \int_{0}^{1} dx     \, {-2\Lambda \over Q}
\left( {\p \over \p x}e^{-xQ/2\Lambda}\right)   \phi(x)\nonumber\\ 
& =& -{2 \Lambda \over Q} e^{-xQ/2\Lambda} \phi(x) \bigg|_{0}^{1} +{2\Lambda \over Q}\int_{0}^{1} dx
\,e^{-xQ/2\Lambda}  {\p  \phi(x) \over \p x}, \nn \\ 
&=& {2 \Lambda \over Q}\phi(0) + ... \nn \end{eqnarray} 
where $x=x_{2}=1-x_{1}$. The last line dropped the exponentially small term from the upper limit.  Using repeated integration by parts gives 
\begin{equation} 
F_{\pi}(Q^{2}) = {2 \Lambda \over Q}\phi(0) + \left({2 \Lambda \over Q}\right)^{2}\phi ^{(1)} (0)+ \left({2 \Lambda \over Q}\right)^{3} \phi^{(2)} (0)+... \label{endexpand} 
\end{equation} 
Here $\phi^{(n)} = \p^{n}\phi/\p x^{(n)}(0)$. It is generally expected that light cone wave functions have non-zero $x$-derivatives at the endpoints. Data from parton distributions supports this. The perception that soft wave function overlaps are incompatible with power-law dependence at large $Q^{2}$ is in error.

The result does not depend strongly on the exponential dependence or factored form of Eq. (\ref{ans}). For example a Gaussian dependence of $\Phi \sim e^{-k_{x}^{2}/\Lambda^{2}} \ra  e^{-x^{2}Q^{2}/\Lambda^{2}}$ uses $\p /\p x^{2}$ integration by parts. Pursuing that, the region $x^{2}Q^{2} \lesssim \Lambda^{2}$ produces just the same power expansion as the region $xQ \lesssim \Lambda$ with different constants. It is also always possible to write \ba \Phi_{\pi}(x, \, k_{x}) = e^{-k_{x}/\Lambda}\tilde \phi(x, \, k_{x}). \nn \ea This simply defines $\tilde \phi(x, \, k_{x})$ as a function with an exponential dependence removed, namely a function that varies more slowly.  Integration by parts proceeds as before, with $ \tilde \phi^{(n)}$ replacing $\phi^{(n)}$. The result is a series in powers of $1/Q$ which is qualitatively unchanged so long as $\tilde \phi$ is slowly varying. 

\subsubsection{Electromagnetic Gauge Invariance} 
\label{sec:gauge}

Phenomenological quark models often violate current conservation. Projecting a $\gamma^{\mu}$ vertex with $g^{\mu \nu} -q^{\mu}q^{\nu}/q^{2}$ hides the problem without actually curing it. The endpoint overlap model with massless quarks satisfies electromagnetic gauge invariance automatically, passing the test $ <P'|q_{\mu}J_{em}^{\mu}|P>=0.$ This is a very detailed topic, explaining why we postponed details of the Dirac algebra in Section \ref{sec:fpi}. 

Let $\ell$ be the difference of quark momenta in one pion. The most general wave function for $J^{P}= 0^{-} \ra fermion, \, anti-fermion$ has Dirac structure  \ba \psi=A \gamma_{5} \slashed P  +B  \gamma_{5}[ \slashed P \slashed \ell - \slashed \ell \slashed P  ] +C \gamma_{5}+D  \gamma_{5} \slashed \ell  . \nn \ea
By permuting $\gamma_{5}$ through the wave function, the terms with one gamma (``chirally even'') have antiparallel helicity and are antisymmetric in spin quantum numbers, and vice versa for chirally-odd. The $A$, $B$ terms are proportional to the large number $P$. These track the Fermion fields which are the largest under a Lorentz boost, 
making them leading order in energy. Thus $A$ is leading in power of $Q$, with zero orbital angular momentum while $B$ is also leading in $Q$, while representing one unit along the momentum axis. (Power counting in $SD$ models is different. Only the $A$ wave function is large as $\ell \ra 0$. ) The $C$, $D$ terms are sub-leading on every basis. 

The form factor's integrand contains a trace over the Dirac indices of wave functions from two pions, three quark propagators, and the perturbative photon vertex ($\gamma_{\mu})$. Chiral selection rules give zero unless even-even and odd-odd chirality wave functions are composed. There are six non-zero combinations denoted $AA', \, BB', \, CC', \, DD', AD',\, BC'$, plus their complex conjugates. Our demonstration of gauge invariance comes from contracting $q_{\mu}$ with the vertex and computing the terms one by one. 

The final result is zero when evaluated using the kinematic conditions of Eq. \ref{sum}. The result is not obvious from manipulating perturbative Ward identities, and the full calculation is quite extensive. The zero-result requires massless quark propagators, namely chirality conservation, as we assume. Apparently a feature of the global chiral symmetry of the model protects the gauge symmetry of the model. We would like to understand this better, while the issue goes beyond the scope of this paper.

\subsubsection{Measuring the Wave Function} 

Experimentally very little is known about the actual large $Q^{2}$ dependence of $F_{\pi}(Q^{2})$. In the space-like region of momentum transfer $Q^{2}F_{\pi}(Q^{2})$ rises and appears to approach a plateau in the region of $Q^{2} \sim few \, GeV^{2}$ \cite{Volmer01,Horn06,Huber08}. That represents a serious problem for $SD$ models, because $Q^{2}$ is far too small for the asymptotic regime to set in. The attempt to interpret the flaw as a ``bonus,'' namely a confirmation of the model's predictions despite contradicting the model's requirements, does not stand up to careful examination. 

\begin{figure}[htbp]
\begin{center}
\includegraphics[width=5in]{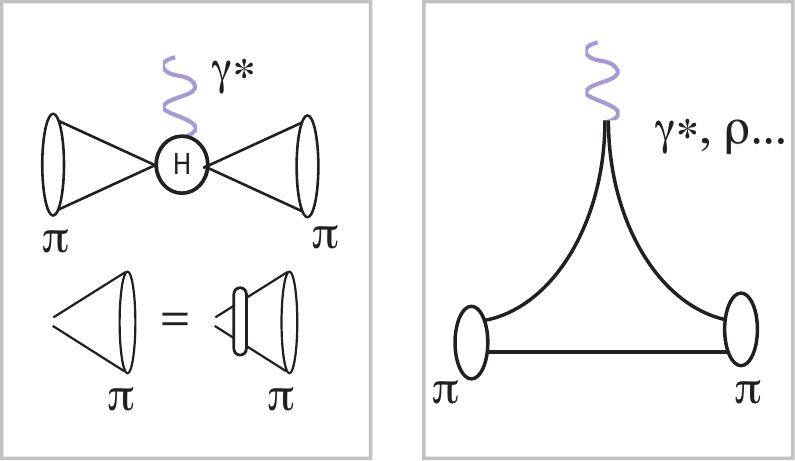}
\caption{The short distance model (left panel) uses approximate pion wave functions developed from summing a subset of diagrams evaluated at the pion pole and with pion quantum numbers. The physics of a $t$-channel $\rho$-meson resonance is absent. The more complete integration regions of the endpoint model (right side) can represent the resonance. }
\label{fig:RhoConcept.pdf}
\end{center}
\end{figure}

It is a serious matter that existing data is well fit by vector meson dominance, associated with the pole at time-like $Q^{2} $ of the $\rho$ meson 
\cite{Sakurai,OConnell,donohue}. 
The importance of the $\rho$ is a problem for $SD$ models. The models 
\cite{Farrar79,LS} are based on expanding the wave function of the pion about zero quark separation. That expansion contains no terms capable of representing the $\rho$ meson's singularity: See Figure \ref{fig:RhoConcept.pdf}.  The asymptotic contribution of the short distance model
\cite{Farrar79} is $F_{\pi SD}= \ab const.  \ab /( Q^{2} log(Q^{2}/\Lambda_{QCD}^{2})$. The normalization $const = 64 \pi^{2}{\cal F}_{\pi}^{2}/9$ with ${\cal F}_{\pi}$=93 MeV is a test of the model. The normalization is too small compared to the experimental data. More ambitious work exploiting self-consistent Bethe-Salpeter wave functions is more convincing \cite{jainmunczek,Alkofer,robertswilliams}. Yet none of the debate about asymptotic limits matters for our approach, which is based on extracting information from what is observable at laboratory energies.

\subsubsection{An Informative Sum Rule}
 
We mentioned in the introduction that folklore about exponentially-dependent form factors is in error. Analysis produces a sum rule for the endpoint-overlap model using Mellin transforms. Represent \ba \Phi(k, x) & = {1 \over 2 \pi i} \int_{-i \infty}^{i \infty} dN \,  \Phi_{N}( x) k^{-N}, \nn \\ & where \quad \Phi_{N}( x)= \int_{0}^{\infty} dk \, k^{N-1}\Phi(k, x) . \nn \ea Positivity of $\Phi$ gives one constraint. The integral over $N$ is done over a contour in a strip where $ \Phi_{N}$ is analytic. Then \ba  F_{\pi} = &{1 \over 2 \pi i} \int_{-i \infty}^{i \infty} dN \, \int_{0}^{\infty} dx \, ( x Q/2)^{-N}\Phi_{N}( x) \nn \\ & = {1 \over 2 \pi i} \int_{-i \infty}^{i \infty} dN \,2^{N}Q^{-N} \Phi_{N, \ 1-N}. \nn \ea 
The right hand side is the Mellin transform of $F_{\pi}$ with respect to $Q$:  
\begin{equation} F_{N} =2^{N} \Phi_{N, \ 1-N}, \nn \end{equation} 
where 
\begin{equation}
\quad F_{N}= \int_{0}^{\infty} dQ\, Q^{N-1}F_{\pi}(Q). \nn \end{equation} 

The formula pinpoints what knowledge of $F$ determines about $\Phi$. The large $Q$ dependence is determined by singularities at $N>0$. These come from any $N_{k}>0$ singularities of the $k$ dependence, and any $N_{x}<1$ singularities of the $x$ dependence. Thus: \bit \im If $\Phi$ falls exponentially with increasing $k$ then $\Phi_{N}( x)$ is analytic for $N>0$. All large $Q$ singularities are found from the $x$ dependence. If $F_{\pi} \sim 1/Q^{2}$, then $\Phi(k \sim 0, \, x)\sim x^{1}$, modulo logarithmic factors. Term by term the $x$ dependence determines the powers and terms in the asymptotic series in $Q^{2}$, while the $k$ dependence contributes to the coefficients.  
\im Suppose $\Phi$ has a power-law tail at large $k$. If $\Phi \sim k^{-2}$ modulo logarithms, then $\Phi_{N}( x)$ will have a pole at $N=2$, producing $Q^{-2}$ dependence. Term by term the $k$ dependence determines the powers and terms in the asymptotic series in $Q^{2}$, while the $x$ dependence contributes to the coefficients, provided $\Phi$ is not as large as $x^{1}$ \im The combination of both $x^{1}$ and $k^{-2}$ can produce a double pole, which translates to possible $Q^{-2}log(Q^{2})$ dependence of $F_{\pi}$. \eit 
The last two options have been extensively explored by the short-distance models. By calculating the large $k$ dependence in the first step, they select in advance {\it one of the regions} the analysis shows are possible in general. The perturbative calculations also agree with the general analysis by producing $Q^{-2}/log(Q^{2})$ behavior from wave functions with corresponding power law behavior. After many years and a considerable investment in manpower, the same calculations have been
found not to apply at finite $Q^{2}$, and also to produce contributions that are too small to explain experimental data \cite{Isgur}. That indicates the endpoint region as being dominant at large $Q^{2}$.

That leaves the first option. Experimentally $Q^{2}F_{\pi}(Q^{2})\sim 0.4 \, GeV^{2}$, which leads to \ba  \Phi_{2, \ -1}=  0.1 \, GeV^{2}, \nn \ea with the leading singularities determined by $\Phi(x)\sim x^{1}$. This contradicts the perturbative $SD$ model wave function (sometimes inappropriately called the ``asymptotic'' wave function). There are no compelling arguments to calculate the $x$-dependence of a wave function with perturbation theory. Ref. \cite{CCRST} fits a numerical calculation to the equivalent of $\Phi(x)\sim x^{1.6}$, which seems to be reasonably consistent. 

\subsubsection{Counting Rule 1}

Summarizing this section, under the universal assumption that $\Phi(x, \,k_{x})$ falls rapidly to constrain $x \lesssim 1/Q$ and $\phi(x)\sim x^{A}$, the cost of overlapping to retain one spectator is a ``slow quark probability'' factor of $1/Q^{A+1}$. 

\subsection{The Proton Electromagnetic Form Factor $F_{1}$ : The Probability of Two Slow Quarks}
 
While our objective focuses on power-counting, we believe value is added by including considerable details in the calculation. Here we compute the proton
electromagnetic form factor, $F_1$, assuming end point domination. 
There have been several earlier calculations of this form factor in different
models \cite{Aznauryan,Duncan80,Avdeenko,Rad84,Ji87,Carlson87,Wagenbrunn,LS1,JK93,BK95,BL99}.  

\begin{figure}[htbp]
\begin{center}
\includegraphics[width=4in]{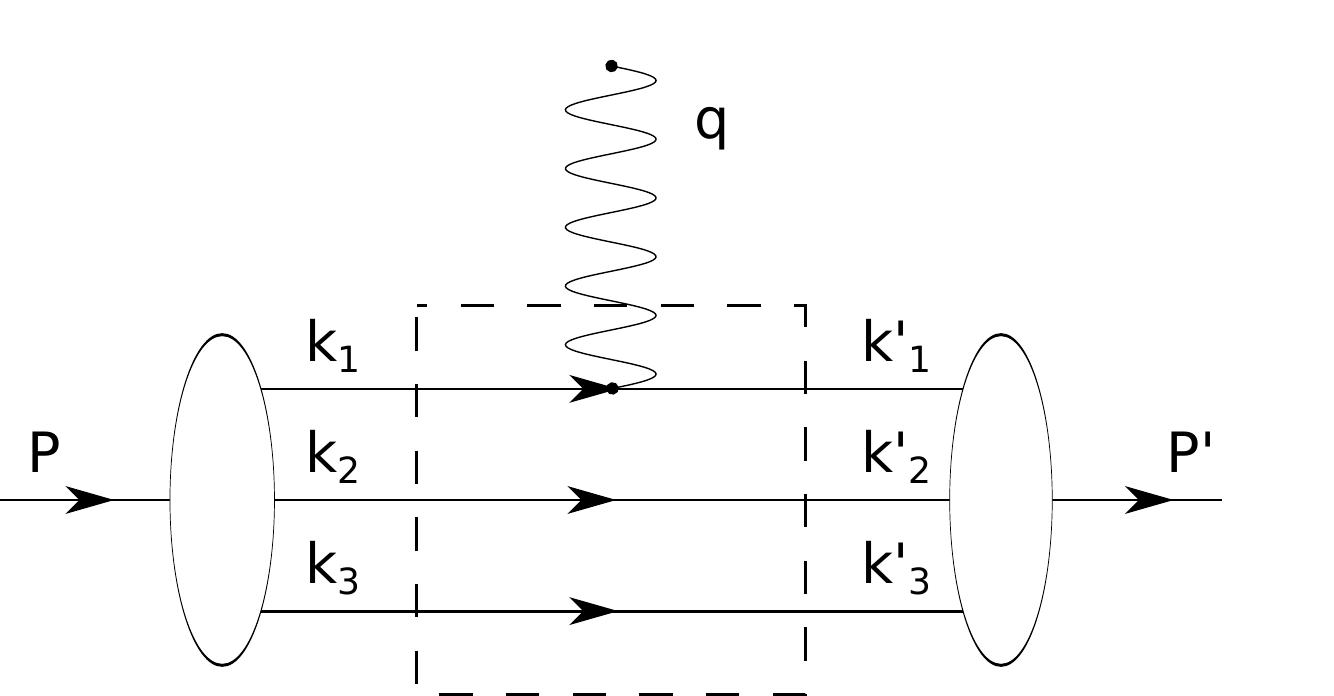}
\caption {\small The proton form factor}
\label{fig:gammapp}
\end{center}
\end{figure}
 
The basic diagram for calculating proton electromagnetic form factors is shown in Fig. \ref{fig:gammapp}. The momenta $P,\,P', \,Q$ are the same as in Eq.(\ref{eq:frame}), and quark momenta use the same notation as Eq. \ref{eq:qmomenta}. Let $Y$ be the proton wave function to three quarks, and let the
electromagnetic vertex be $\Gamma^{\mu} =-ie\gamma^{\mu}\,\delta^{4}(k_{1}+Q-k'_{1})\,\delta^{4}(k_{2}-k'_{2})\,\delta^{4}(k_{3}-k'_{3})$. The matrix element for the process is
\begin{equation}
\int \prod_{i}\frac{d^{4}k_{i}}{(2\pi)^{4}} \frac{d^{4}k'_{i}}{(2\pi)^{4}} ( \overline{Y^{\prime}}(k'_{i}) \times \Gamma^{\mu}\times Y(k_{i}) ) = -i eF_{1}(Q^{2})(\overline{N}^{\prime}\gamma^{\mu}N) + F_{2}(Q^{2})(\overline{N}^{\prime}\iu\sigma^{\mu\nu}q_{\nu}N),  \label{formy}
\end{equation}
where $N, \, \overline{N}^{\prime}$ are Dirac spinor functions. 
The momentum space wave functions with leading power of $P$ is written as \cite{CZ,Ioffe}
 \begin{equation}      Y_{\alpha\beta\gamma}(k_{i},P) = \frac{f_{N}}{16\sqrt{2}N_{c}}\{ 
     (\slashed{P}C)_{\alpha\beta}(\gamma_{5}N)_{\gamma}\mathcal{V} + (\slashed{P}\gamma_{5}C)_{\alpha\beta}N_{\gamma}\mathcal{A} + \iu (\sigma_{\mu\nu}P^{\nu}C)_{\alpha\beta}
 (\gamma^{\mu}\gamma_{5}N)_{\gamma}\mathcal{T}\}.\label{eq:lipwavef}
 \end{equation}

 Here $\alpha,\beta,\gamma$ are Dirac indices, $\mathcal{V,A,T}$ are scalar functions of the quark momenta($k_{i}$), 
        $N_{c}$ is the number of colors,
        $C$ the charge conjugation operator, 
        $\sigma_{\mu\nu}= \frac{\iu}{2}[\gamma_{\mu},\gamma_{\nu}]$,
and $f_{N}$ is a normalization.

\begin{figure}[htbp]
\begin{center}
\includegraphics[width=3in]{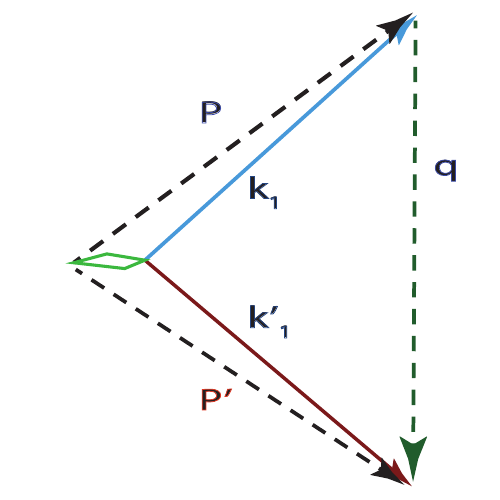}
\caption{ \small Endpoint kinematics in the proton case. Proton momenta are shown as dashed lines, quark momenta as solid lines. Isosceles triangles representing energy and momentum conservation close as in Figure \ref{fig:EndPion.pdf}. The momenta of all spectator constituents sum to cover the differences $P-k_{1}$ and $P'-k'_{1}$. }
\label{fig:EndpointKineProton.pdf}
\end{center}
\end{figure}

Several combinations appearing in $\overline{Y^{\prime}}\Gamma^{\mu}Y$ give leading order contributions to $F_{1}$. Consider, for example, 
\begin{equation}       (\overline{N}^{\prime}\gamma_{5}\gamma^{\rho^{\prime}})_{\gamma^{\prime}}\gamma^{\mu}_{\gamma^{\prime}\gamma}(\gamma^{\rho}\gamma_{5}N)_{\gamma} 
  = (\overline{N}^{\prime}\gamma^{\mu}N)2g^{\rho^{\prime}\rho} + \dots \label{item}
\end{equation}
Collecting all coefficients proportional to this term written gives
\begin{equation}
         -(C^{-1}\slashed{P}^{\prime})_{\alpha\beta}(\slashed{P}C)_{\alpha\beta}\mathcal{V'}\mathcal{V}
          + (C^{-1}\gamma_{5}\slashed{P}^{\prime})_{\alpha\beta}(\slashed{P}\gamma_{5}C)_{\alpha\beta}\mathcal{A'}\mathcal{A}
          - 2g^{ab}(C^{-1}\sigma_{\nu a}P^{\prime\nu})_{\alpha\beta}(\sigma_{b\nu}P^{\nu}C)_{\alpha\beta}\mathcal{T'}\mathcal{T} \label{another}
\end{equation} 
Inserting Eq. \ref{another} in Eq. \ref{formy} gives 
\begin{equation}   
F_1(Q^2)\sim \int [dxdk_{T}][dx'dk'_{T}] \Psi_{1tot}\delta^{4}(k_{1}+q-k'_{1})\delta^{4}(k_{2}-k'_{2})\delta^{4}(P+q-P')
    \label{eq:int}
\end{equation}
where,
\begin{equation}
 \Psi_{1tot}= \left(\frac{f_{N}}{16\sqrt{2}N_{c}}\right)^{2}\left(\frac{Q}{2}\right)^{2}\{ -8\mathcal{V'}\mathcal{V} 
         - 8 \mathcal{A'}\mathcal{A} + 48\mathcal{T'}\mathcal{T} \}\nn
\end{equation}
and 
\begin{equation}
[dxdk_{T}] =  dx_{1}d^{2}\vec{k}_{1}dx_{2}d^{2}\vec{k}_{2}dx_{3}d^{2}\vec{k}_{3}\delta(x_{1}+x_{2}+x_{3}-1)\delta^{2}(\vec{k}_{1}+\vec{k}_{2}+\vec{k}_{3})
\end{equation}
The delta functions lead to kinematics similar to Eq.(\ref{sum}). Momentum conservation requires $k_{1}^{\mu}=k'^{\mu}_{1}+q^{\mu}$ and $k_{2}=k'_{2}$. One transverse momentum of the struck quark is unconstrained except by wave functions, while the other transverse momentum is constrained by the relation previously found for the pion: \ba -x_{1}Q/2-k_{1x}+Q & = x'_{1}Q/2+k'_{1x} \label{protonsum}; \\
 k_{1x}  & =  {(1-x_{1})Q\over 2}={(x_{2} +x_{3})Q\over 2}. \nn \ea 
The measure is then replaced by
\ba & [dxdk_{T}][dx'dk'_{T}] \rightarrow dx_{1}dx_{2}dk_{y1}dk_{y2}\frac{1}{Q^{2}}.\label{eq
:endpoint} \ea
Integrating over the unconstrained variables gives \ba \int  dk_{y1}dk_{y2} \Psi_{1tot} = \Phi_{1P}(\tilde k_{1x}(x_{1}), \, x_{1},
\, x_{2}). \nn \ea That leaves the integration depending on $Q$ as \ba F_{1} = \int\limits_{0}^{1} d x_{1} dx_{2} \, \Phi_{1P}(k_{1}=(1-x_{1})Q / 2 
, \, x_{1},x_{2}). \nn \ea Once more consider an exponential ansatz 
\ba  
    \mathcal{V',A',T'}&\propto (1-x'_{1})x'_{1}\ \psi(\vec k'_{T})\nonumber\\ 
    \mathcal{V,A,T} &\propto (1-x_{1})x_{1}\ \psi(\vec k_{T})
    \label{eq:wavefunc}
\ea
where $ \psi(\vec k_T) \sim \exp\left[-\left(\vec k_T\right)^2/\Lambda^2\right]$.  That leads to
\ba \Phi_{1P}(k_{1}, x_{1},x_{2})\sim e^{- k^{2}_{1x}/\Lambda^{2}} \phi(x_{1},x_{2}). \nn \ea
Evaluated at $k_{1x}=(x_{2} +x_{3})Q / 2$, both $x_{2}$ and $x_{3}$ range over intervals of size $\Lambda/Q$. For wave functions that
are uncorrelated products, $\psi = \psi(x_{1}) \psi(x_{2})$, the probability of finding two slow quarks is precisely the product of two
slow quark probabilities. For quark wave functions going like $x_{1}^{A}(1-x_{1})^{A}$  we find \ba \quad \Phi_{1P} \sim&  x_{2}^{A} x_{3}^{A}+\dots; \nn \\ &  F_{1} \sim 1/Q^{4A}.  \nn \ea Experimental data finds that $Q^{4}F_{1} \sim constant$ for $Q^{2} \geq GeV^{2}$. The data indicates that $A\sim 1$, namely that quark wave functions must go like $x_{i}^{1}$ near $x_{i}\sim 0$. We emphasize that this result does not require extremely large $Q^{2}$ (large logarithms of $Q^{2}$). The estimates are based on comparison with the transverse size of the proton, $Q^{2} >> \Lambda^{2}$ for $\Lambda^{2} \sim 0.1 \, GeV^{2}$. 

\subsubsection{A Typical Perturbative Question}

The approach based on short-distance perturbation theory typically asks how these relations behave when ``soft gluons'' are added. 

The notion of adding soft gluons is tied to a basis of Fock state wave functions used in perturbation theory. Since the interacting theory is not the free theory, Feynman diagrams represent the interactions with gluons of all momenta. That does not represent our model: By construction, we are concerned with the full wave functions of the interacting theory. Thus the calculation using the fully interacting wave functions is self-consistent {\it without} adding gluons. 

The question of soft gluons is vital for the internal consistency of the $SD$ model and its (conceptually different) estimate of the endpoint contribution. {\it Assuming one desires} a perturbative description, experience indicates that any finite number of soft gluon internal diagrams will not revise the leading power of $Q^{2}$. Considerable effort has gone into arranging calculations that would be simultaneously compatible with the assumptions of short distance. 
That has led to statements that Sudakov effects suppress the endpoint region. These statements refer to the short-distance model, not ours. If it is true that the endpoint region of the $SD$ model is negligible, it has no bearing on the endpoint region of all possible models expressed in different quantum-mechanical bases. None of it is our concern once the focus is on extracting non-perturbative information from experiments. 

Nevertheless it is interesting to check that soft-gluons do not change the leading power behavior. We feel that a specific calculation is more convincing than an estimate, and present one in the Appendix. 

\subsubsection{The Pauli Form Factor $F_{2}$}

We have investigated the large $Q^{2}$ dependence of the Pauli form factor $F_{2}$ in the endpoint model. We find $1/Q^{5}$ dependence occurs in more than one way, {\it together with} wave functions that go like $x(1-x)$. This result is surprising and impinges directly on the issue of quark orbital angular momentum\cite{oam} and ``the shape of the proton'' \cite{Miller03,Gross06}. Yet the calculations we have available are complicated, and too detailed to be appropriate to review here. We
plan to present them in an future paper\cite{F2Sumeet}. For the purposes of this survey, we can objectively report that {\it it has not been shown} that experimental data \cite{Jones00} measuring $F_{2}$ at large $Q^{2}$ is in conflict with power-counting of the endpoint overlap model. The lack of previous work is itself remarkable because {\it it has been shown} that $F_{2}\sim 1/Q^{5}$ is incompatible with short distance models.

\subsection{Endpoint Constituent Counting Rules for Form Factors} 

We are now in a position to state the leading power endpoint constituent counting rules for form factors. Let there be $n_{IN}$ ($n_{OUT}$) quarks in the $IN$ ($OUT$) state hadron.  For now choose $n_{IN}=n_{OUT}=n$. One constituent is scattered, requiring $n-1$ quarks to be slow. The form factor is given by induction, \ba F(n \, quarks) = \int d x_{1} dx_{2}...dx_{n-1} \, \Phi_{1P}(\tilde k =\sum_{j=1}^{n-1}x_{j }Q / 2 , \, \sum_{j=1}^{n-1}x_{j}  ). \nn \ea For wave functions with linear dependence as $x_{j} \ra 0$, each extra constituent beyond the valence configuration causes a suppression of the form factor by a power of $1/Q^{2}$. The leading power dependence is \ba F(n \, quarks) \sim {1\over (Q^{2})^{n-1}}. \nn \ea These happen to be the same rules as the early ``dimensional'' counting rules\cite{GRF}, but for entirely different reasons. Hard propagator factors are not the explanation. The explanation lies in the phase space to find quarks available to scatter. The general dominance of phase space over hard scattering is reminiscent of the independent scattering mechanism originally discovered by Landshoff\cite{Landshoff,Donnachie84}.

We will be straightforward with what is new in the power law. It has been noticed again and again that the endpoint contribution cropped up and competed with the short-distance model of form factors. Many papers have dealt with the issue as a troublesome instability of short-distance dominance. Yet we are not aware of an explicit, {\it positive} statement of the predictive regularity between the number of scattered constituents and the observed power laws. We cannot explain why the universal potential of endpoint overlap models has not been not widely recognized. 

\section{Hadron-Hadron Exclusive Reactions}
\label{sec:hadrons}

In this Section we find that the experimentally observed power laws of hadron-hadron exclusive reactions \cite{Landshoff_Pol,BL80,Fiore99,Asad84,Quiros81,Stone78,Chen77,Islam75,Fishbane73,Collins74,Akerlof66,Allaby67,Carrigan70,Cocconi65,Pire82} are explained by the endpoint overlap model. Unlike the $SD$ model, no approximations of an asymptotic character are needed. The approximations assume only that $Q^{2} >>\Lambda^{2}$. 

\subsection{General Features}

We are not aware of a previous focused effort to study the contributions of the endpoint overlap model in $2 \ra 2$ hadron-hadron reactions. The power counting for both the $SD$ and endpoint models are expedited by a simple observation. The amplitude of almost all contributions scales like the combination of form factor amplitudes. This observation is quite old\cite{yang}, and developed for a different purpose, yet it is rather general. If there is another contribution with a qualitatively different behavior, its momentum flow will go by a qualitatively different topology. 

In the high energy limit the differential cross section for $2 \ra 2$ reactions with amplitude $M$ is given by \ba {d \sigma \over dt } = const. {MM^{*} \over s^{2}}. \nn \ea The composition of two form factors $F_{A}F_{B}$ with a $1/Q^{2}$ exchange kernel scales like $ F_{A}F_{B}Q^{2}/Q^{2}$. The numerator factor of $Q^{2}$ accounts for the vector vertex factors not contained in $F_{i}$. 

There is one significant difference between models, however. Multiple gluon exchanges in the $SD$ model have no strong selection rules 
from the color singlet nature of hadrons. Scattering a single constituent in the endpoint overlap model requires at least two gluons in a singlet combination. It is well known that the box diagram of two gluon exchanges scales with just the same power of $Q^{2}$ as a single gluon exchange, times logarithmic factors that are exactly computable\cite{stavros}. For our purposes multi-gluon exchanges are indistinguishable, and at most laboratory momentum transfers, probably necessary. 

To proceed: For {\it fixed angle} kinematics $s\sim t$, the endpoint overlap model composing form factors predicts: \bit \im For $\pi \pi \ra \pi \pi $, \ba M \sim 1/Q^{4} \sim 1/s^{2}; \nn \\ {d \sigma \over dt } \sim {1/s^{6}}. \nn \ea  \im  For $\pi p \ra \pi p $, \ba M \sim 1/Q^{6} \sim 1/s^{3}; \nn \\ {d \sigma \over dt } \sim {1/s^{8}}. \nn \ea  For $ p p \ra p p $, \ba M \sim 1/Q^{8} \sim 1/s^{4}; \nn \\ {d \sigma \over dt } \sim {1/s^{10}}. \nn \ea \im For $2 \ra 2$ scattering of
hadrons with $n_{1}$ and $n_{2}$ valence constituents, \ba M \sim 1/Q^{n_{1}+n_{2}+2} \sim 1/s^{(n_{1}+n_{2})/2+1}; \nn \\ {d \sigma \over dt } \sim {1/s^{n_{1}+n_{2}+ 4}}. \nn  \ea \eit Agreement with experiment 
\cite{Landshoff_Pol,BL80,Asad84,Stone78,Akerlof66,Allaby67,Carrigan70,Cocconi65,Jenkins:1978rm,Stone:1977ik} are explained by the endpoint overlap model. Unlike the $SD$ model, no approximations of an asymptotic character are needed. The approximations assume only that $Q^{2} >>\Lambda^{2}$, 
 adds support to the valence state of the pion having two constituents, and the proton having three. As before, scattering constituents beyond the valence components is suppressed by powers of $1/s$.

The counting is different for the $t$ dependence of amplitudes at fixed $s>> GeV^{2}$. In that case $MM^{*}/s^{2} \sim MM^{*}$. With $|t|<<s$ the leading dependence replaces $s \ra t$, and multiplies the results above by $t^{2}$. By far the most important example comes from $pp \ra pp$ scattering, which displays a stunning experimental dependence falling like $t^{-8}$ \cite{Landshoff,Collins75,Donnachie79,Collins82}: 
exactly the endpoint-overlap contribution. 

\subsubsection{Discussion}

We mentioned that a qualitatively different momentum flow could change the counting. The independent scattering model\cite{Landshoff} is usually highlighted to explain the $t^{-8}$ dependence. Landshoff had earlier found the model by not making the same assumptions of the model of Brodsky and Farrar. The independent scattering ($IS$) model gets its power law partly from the phase space of fast quarks with $x\sim 1/3$ to overlap with the wave function in the final state. There are three (3) hard vector exchanges, suppressing the amplitude by corresponding powers of $1/t$. In comparison the endpoint overlap contribution uses one (1) hard vector exchange, while obtaining the same powers of $1/t$ from the probability to find two quarks near the endpoint. 

The phenomenology of complex phases and spin dependence are very similar for the $IS$ and endpoint models. When treated in Fock-basis perturbation theory both model have similar Sudakov factors. Such factors may well explain the oscillations seen in $pp$ fixed angle scattering and color transparency. Based on the results of Mueller\cite{muellersaddle}, we conjecture that saddle point interpolation between the endpoint model and the $SD$ model will be the dominant asymptotic amplitude. This is because the Sudakov suppression of one fast quark is less severe than the three fast quarks of the $IS$ model. 

\subsection{Supporting Calculation} 

As with the form factors, we believe that supporting calculations are at least as important as general arguments. 

\begin{figure}[!t]
\centering
\includegraphics[scale=0.50,angle=0]{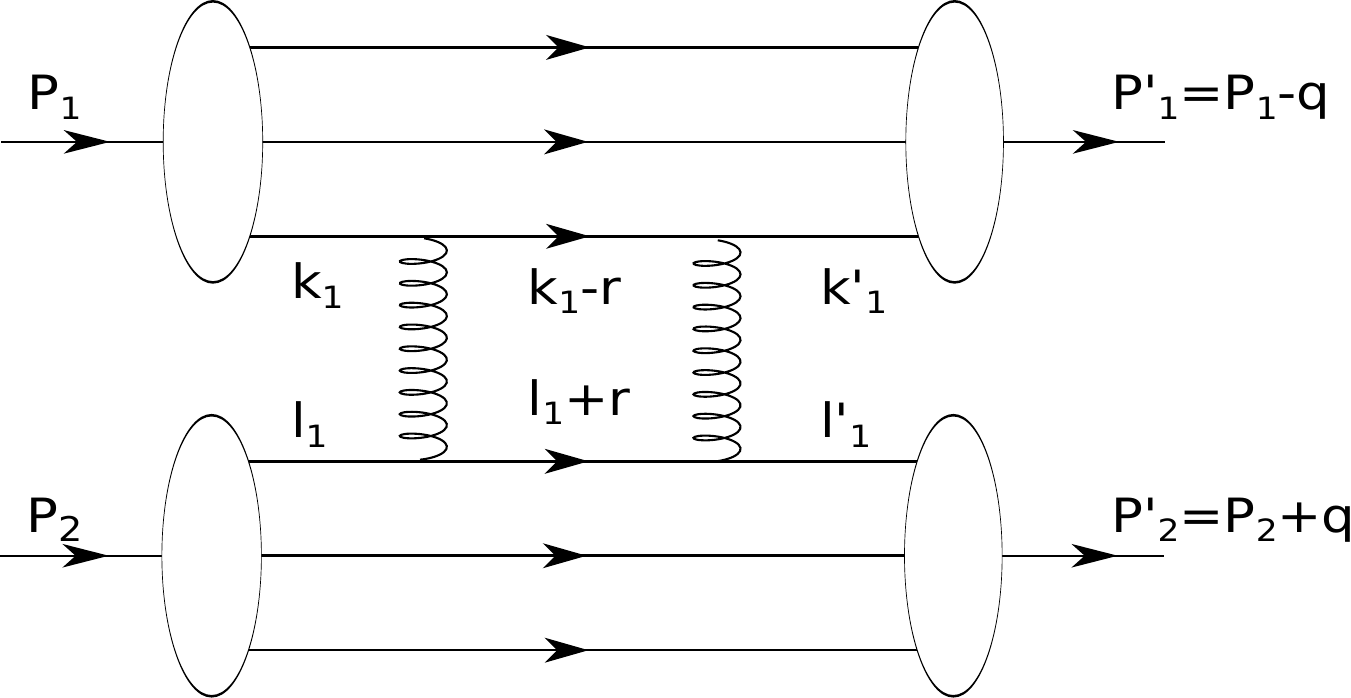}
\caption {\small Proton-proton elastic scattering with a minimal hard-scattering kernel.}
\label{fig:QCD}
\end{figure}

Consider proton-proton scattering of Fig.\ref{fig:QCD}, $\, \bf{p}(P_{1})+\bf{p}(P_{2}) \ra \bf{p}(P'_{1})+\bf{p}(P'_{2})$ in the limit $s=(P_{1}
+P_{2})^{2}\sim t =(P'_{1}-P_{1})^{2}=q^{2}=-Q^{2}$. In the center of mass frame the momenta of the two incoming particles are \ba 
    P_{1}  =&  (p,0,0,-p), \nn \\ 
   P_{2}  = & (p,0,0,p). \nn \ea  
The amplitude for the scattering diagram is given by
\begin{equation}
    \begin{split}
        M \propto & \left(\frac{-\iu g_{s}}{2}\right)^4 \int \frac{d^{4}r}{(2\pi)^{4}} 
         \frac{-\iu g_{\mu_{1}\nu_{1}}}{(q-r)^2}   \frac{-\iu g_{\mu_{2}\nu_{2}}}{r^2}       \\
         & \int \prod_{i}\frac{d^{4}k_{i}}{(2\pi)^{4}}\frac{d^{4}k'_{i}}{(2\pi)^{4}} 
         \overline{Y}_{\alpha^{\prime}\beta^{\prime}\gamma^{\prime}}(k'_{i},P'_1)\left[\gamma^{\mu_{1}}\frac{\slashed{k}_{1}-\slashed{r}+m}
{(k_{1}-r)^2-m^{2}+\iu\epsilon}\gamma^{\mu_{2}}\right]_{\gamma^{\prime}\gamma}\delta_{\alpha^{\prime}\alpha}\delta_{\beta^{\prime}\beta}Y_{\alpha\beta\gamma}(k_{i},P_{1})\\
&   \int \prod_{j}\frac{d^{4}l_{j}}{(2\pi)^{4}} \frac{d^{4}l'_{j}}{(2\pi)^{4}}
\overline{Y}_{\alpha^{\prime}\beta^{\prime}\gamma^{\prime}}(l'_{i},P'_{2})\left[\gamma^{\nu_{1}}\frac{\slashed{l}_{1}+\slashed{r}+m}
{(l_{1}+r)^2-m^{2}+\iu\epsilon}\gamma^{\nu_{2}}\right]_{\gamma^{\prime}\gamma}\delta_{\alpha^{\prime}\alpha}\delta_{\beta^{\prime}\beta}Y_{\alpha\beta\gamma}(l_{i},P_{2})
    \end{split}
\end{equation}
As with the case of proton form factor, the interaction vertex will have delta functions enforcing the conservation of momentum
in the quark interactions, which are implicit in the above expression.

Extract the integral over the free momentum $r$, given by 
\begin{equation}
        \int \frac{d^{4}r}{(2\pi)^{4}} \left(\gamma^{\mu_{1}}\frac{\slashed{k}_{1}-\slashed{r}+m}
{(k_{1}-r)^2-m^{2}+\iu\epsilon}\gamma^{\mu_{2}}\right)
\frac{1}{(q-r)^2}\frac{1}{r^2}
\left(\gamma^{\nu_{1}}\frac{\slashed{l}_{1}+\slashed{r}+m}
{(l_{1}+r)^2-m^{2}+\iu\epsilon}\gamma^{\nu_{2}}\right)
\end{equation}
Integration is performed using Feynman parametrization. Simplifying the denominator using Feynman parameters$(a_{i})$ leads to a denominator $D$ given by,
\begin{equation}
        D = (l^{2}-\Delta+\iu\epsilon)^{4}
\end{equation}
where \ba &  l = r - a_{1}k_{1} - a_{3}q + a_{4}l_{1}; \nn \\ & \Delta = [ a_{1}k_{1} + a_{3}q - a_{4}l_{1}]^{2} - a_{1}k_{1}^{2}-a_{3}q^{2}-a_{4}l_{1}^{2}+a_{1}m^{2}-a_{4}m^{2}+\iu\epsilon. \nn \ea  
We may neglect terms of the form $k_{1}^{2}, \, l_{1}^{2}$ assuming the quarks are nearly light like.
Terms of the form $k_{1}\cdot q, l_{1}\cdot q, k_{1}\cdot l_{1}$ are of the same order as $Q^{2}$, assuming $t \sim s$. Thus the dominant contribution in the $\Delta$ goes like $Q^{2}$.

Terms in the numerator of the form $l^{\mu}k_{1}^{\nu}, \, l^{\mu}q^{\nu}, \, l^{\mu}l_{1}^{\nu}\dots$ vanish upon integration. The other terms can be integrated using the standard substitution
\begin{equation}
        l^{\mu}l^{\nu}\text{term}\Rightarrow \frac{g^{\mu\nu}}{\Delta} 
    \propto \frac{g^{\mu\nu}}{Q^{2}} \hspace{10mm}
       \end{equation} In comparison, the other terms in the numerator scale like $1/Q^{4}$. 
To leading power we keep the $l^{\mu}l^{\nu}$ term. The amplitude is given by
\ba  & -\left(\frac{g_{s}}{2}\right)^{4}\frac{1}{Q^2}
\int \,  [da_{i} ]\nn \\
 &     \times    \int \prod_{i}\frac{d^{4}k_{i}}{(2\pi)^{4}} \frac{d^{4}k'_{i}}{(2\pi)^{4}} 
\overline{Y}_{\alpha^{\prime}\beta^{\prime}\gamma^{\prime}}(k'_{i},P'_1)[\gamma^{\mu_{1}}\gamma^{\mu}\gamma^{\mu_{2}}]_{\gamma^{\prime}\gamma}\delta_{\alpha^{\prime}\alpha}\delta_{\beta^{\prime}\beta}Y_{\alpha\beta\gamma}(k_{i},P_{1})\\
& \times \int \prod_{j}\frac{d^{4}l_{j}}{(2\pi)^{4}}\frac{d^{4}l'_{j}}{(2\pi)^{4}}
\overline{Y}_{\alpha^{\prime}\beta^{\prime}\gamma^{\prime}}(l'_{i},P'_2)[\gamma_{\mu_{1}}\gamma_{\mu}\gamma_{\mu_{2}}]_{\gamma^{\prime}\gamma}\delta_{\alpha^{\prime}\alpha}\delta_{\beta^{\prime}\beta}Y_{\alpha\beta\gamma}(l_{i},P_{2}). \nn
\ea Here \ba [da_{i}] =\prod_{i}^{4}da_{i} \delta(\sum_{j}^{4} a_{j}-1). \nn \ea 
The calculation can be simplified by a Lorentz transformation to a frame where the momenta of the protons becomes equivalent to the
momenta of Eq. (\ref{eq:frame}). 
For example
$$ k_{1}+k_{2}+k_{3} = P_{1} = (p,0,0,-p) \xrightarrow{\text{\tiny{Lorentz transform}}} k^{L}_{1}+k^{L}_{2}+k^{L}_{2} = P^{L}_{1}=(Q/\sqrt{2}, \, -Q/2, \, 0, \,Q/2 )$$
\begin{equation}\begin{split}\nn k'_{1}+k'_{2}+k'_{3} = P_{1}+q =& (p-q_{0},-q_{1},-q_{2},-p-q_{3}) \\
&\xrightarrow{\text{\tiny{Lorentz transform}}} k^{'L}_{1}+k^{'L}_{2}+k^{'L}_{2} = (P_{1} - q)^{L}=(Q/\sqrt{2}, \, +Q/2, \, 0, \,Q/2 )
\end{split}\end{equation}
Such a transformation will allow the use of results of the proton form factor calculation.

Substituting the wave function from Eq. (\ref{eq:lipwavef}), a single term from the wave function is sufficient
to understand the behavior of this integral. We illustrate the term $M_{V}$ going like $(\slashed{P}C)_{\alpha\beta}(\gamma_{5}N)_{\gamma}\mathcal{V}$, which is 
\begin{equation}
    \label{eq:awithwf}
    \begin{split}
       M_{V}= & (-\frac{g_{s}}{2})^{4} \frac{1}{Q^2}\int \prod_{i}^{4} [da_{i}] \\ &\int
        [dxdk_{T}][dx'dk'_{T}](\slashed{P_{1}}C)_{\alpha\beta}(C^{-1}(\slashed{P_{1}}-\slashed{Q}))_{\alpha\beta} 
 [\overline{N}_{P_{1}-Q}\gamma^{\mu_{1}}\gamma^{\mu}\gamma^{\mu_{2}}N_{P_{1}}]\mathcal{V}(k_{i},P_{1}) 
 \mathcal{V'}(k'_{i},P_{1}-Q)\\ & \times\int [dydl_{T}][dy'dl'_{T}] (\slashed{P_{2}}C)_{\alpha\beta}(C^{-1}(\slashed{P_{2}}+\slashed{Q}))_{\alpha\beta} 
 [\overline{N}_{P_{2}+Q}\gamma_{\mu_{1}}\gamma_{\mu}\gamma_{\mu_{2}}N_{P_{2}}]\mathcal{V}(l_{i},P_{2})\mathcal{V'}(l'_{i},P_{2}+Q) 
    \end{split}
\end{equation}
From the calculations following Eq.(\ref{eq:int}), the integrations become
\begin{equation}
    (\slashed{P_{1}}C)_{\alpha\beta}(C^{-1}(\slashed{P_{1}}-\slashed{Q}))_{\alpha\beta} 
    \int[dxdk_{T}][dx'dk'_{T}]\mathcal{V}(k_{i},P_{1})\mathcal{V}^{\prime}(k'_{i},P_{1}-Q) \propto \frac{1}{(Q)^4} \nn 
\end{equation}
\begin{equation}
    (\slashed{P_{2}}C)_{\alpha\beta}(C^{-1}(\slashed{P_{2}}+\slashed{Q}))_{\alpha\beta} 
    \int[dydl_{T}][dy'dl'_{T}]\mathcal{V}(l_{i},P_{2})\mathcal{V}(l'_{i},P_{2}+Q) \propto \frac{1}{(Q)^4} \nn
\end{equation}
That implies
\begin{equation}
   M \propto  -\left(\frac{g_{s}}{2}\right)^{4}\left(\frac{1}{Q^2}\right)
    \int \{da_{i}\} \frac{1}{(Q)^4} [\overline{N}_{P_{1}-Q}\gamma^{\mu_{1}}\gamma^{\mu}\gamma^{\mu_{2}}N_{P_{1}}]   
\frac{1}{(Q)^4} [\overline{N}_{P_{2}+Q}\gamma_{\mu_{1}}\gamma_{\mu}\gamma_{\mu_{2}}N_{P_{2}}]  \nn 
\end{equation}
Calculate the cross section using
\begin{equation}
    \begin{split}
        {d\sigma\over dt} \sim {|{\cal M}|^2\over s^2} \propto \frac{1}{s^2}\left(\frac{1}{(Q)^4}\right)^{4}\left(\frac{1}{Q^2}\right)^{2}&Tr[(\slashed{P_{1}}-\slashed{Q})\gamma^{\mu_{1}}
\gamma^{\mu}\gamma^{\mu_{2}}\slashed{P}_{1}\gamma^{\nu_{1}}\gamma^{\nu}\gamma^{\nu_{2}}]\\
    &  Tr[(\slashed{P_{2}}+\slashed{Q})\gamma_{\mu_{1}} \gamma_{\mu}\gamma_{\mu_{2}}\slashed{P}_{2}\gamma_{\nu_{1}}\gamma_{\nu}\gamma_{\nu_{2}}]
    \end{split} \nn 
\end{equation}
The leading term from simplifying the trace goes like $p^{4}$. Using $s \sim t$ we find
\begin{equation}
    {d\sigma\over dt} \propto \frac{1}{s^2}\frac{1}{(Q)^{16}}\frac{1}{Q^4} p^{4}\propto \frac{1}{s^{10}} \nn 
\end{equation} Many other terms give a similar $s$ dependence.

\section{Concluding Remarks}

\label{sec:conclude}

We have shown that the endpoint overlap model stands as a comprehensive theory of hadronic reactions at large momentum transfer. It explains the observed experimental regularities in all cases we have investigated. The history of endpoint dominance is curious, and possibly explains why the model failed to be completely developed. 

In 1970 Drell and Yan \cite{Drell70}, and later West \cite{West70}  ($DYW$) discussed a partonic model connecting hadronic form factors to deeply inelastic scattering. Using symbol $\eta$ for the parton momentum fraction since called $x$, the central region $1-\eta \geq \Lambda/Q$ was found to predict a form factor falling too fast to agree with data. From this region $F_{1}(Q^{2}) \sim g(Q^{2})/Q^{2}$, with $g(Q^{2}) \sim exp(-Q^{2}/\Lambda^{2})$ is expected. In comparison, the endpoint
region $1-\eta \leq \Lambda/Q$ was observed to predict $F_{1}(Q^{2}) \sim (1/Q^{2})^{(p+1)/2}$, if the structure function $\nu W_{2} \sim (1-\eta)^{p}$. The value $p=1$ was computed by Drell and Yan \cite{Drell70} in a prototype {\it two constituent} calculation. The value $p=1$ was also noted as being too small to fit the data needing $p \geq 2$. While the two components of the toy model were a pion and a nucleon, the 1970 calculation in our view constitutes the prototype ``constituent counting'' relation connecting the scaling power with the number of constituents. The paper appears approximately five years before the papers of Brodsky and Farrar \cite{GRF} and Matveev {\it et al.} \cite{matveev}, which found counting rules on a different basis.

Subsequently many workers noticed that a calculation of the endpoint contribution would be revised by a power of $1/Q^{2}$ for each additional constituent added to a given process. Many workers also concluded that the endpoint contribution might be dominating the calculation of their particular process.  Yet the endpoint region never saw anything like the degree of development of the short-distance perturbative model. For reasons we cannot explain, we cannot find a reference strongly advocating for the endpoint region, and developing it as a ``comprehensive theoretical picture'' that explains the observed power law dependence. 

There exists a possible explanation coming from the drive of the early era. That time was concerned with testing the Lagrangian of QCD, and without needing to know wave functions. Exclusive reactions were hardly a good testing ground.  Imagine trying to test perturbative quantum electrodynamics in a world where the Hydrogen atom bound states had not been solved. In such a Universe, calculations actually depending on unknown wave functions would be ``bad.'' That is, they would be bad for establishing a Lagrangian by perturbation theory. Calculations for Hydrogen-Hydrogen scattering not depending on unknown wave functions would be very difficult to concoct. However, once any scheme self-consistent with perturbation theory was found, it would be ``good'' whether or not it was incomplete and ``wrong.'' 

We believe that the lure of a strictly perturbative procedure caused a false perception that ``correct physics'' could only depend on operator product expansion moments of wave functions. The operator product expansion is an ansatz of great power when it applies, while creating huge gaps when it does not, but this was not obvious right away. Once the attitude was adopted, the opportunity to use data to learn about wave functions was rejected. As a result, the opportunity to actively use data to learn about hadron structure remains a relatively unexplored field.
It has taken 30 years of more and more detailed calculations to find that the tiny integration region of the short-distance model can at most be relevant for extremely large $Q^{2}$ which need momentum transfers that are tens, hundreds, or thousands of times larger than laboratory scales to be relevant. For emphasis, we do not know of a single calculation that strongly supports the numerical dominance of the short-distance region, relentless advocacy notwithstanding. Indeed the first step towards arranging for short distance dominance has been to banish the endpoint region as perturbatively inconsistent, which (we maintain) {\it is a signal of a concept error}. 

Through the entire period the endpoint contribution has never gone away. The time has come to accept endpoint contribution, and explore it further.  \\

{\bf {Acknowledgments:}} Some of this work was initiated during a University of Washington-INT workshop. We thank G. A. Miller and Toly Radyushkin for discussions.  We also thank Leonard Gamberg, Ron Gilman, Simonetta Liuti, Zein Eddine Meziani, Charles Perdrisat and Oscar Rondon for insights. We also thank 
Dipankar Chakrabarti for collaborating at the initial stages of this 
work.

\section{Appendix: Soft Gluon Effects}

\begin{figure}[!t]
\centering
\includegraphics[scale=0.50,angle=0]{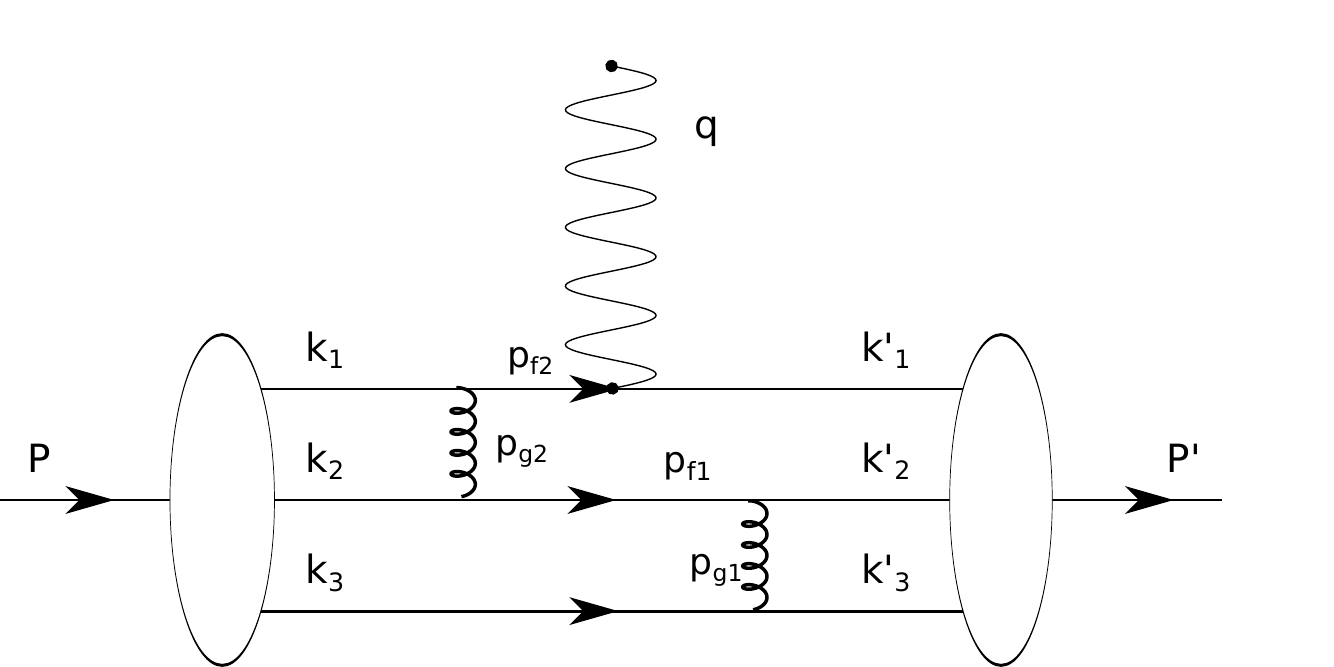}
\caption {\small A 2 gluon exchange contribution to the proton form factor}
\label{fig:2gluon}
\end{figure}

Soft gluon effects are an intrinsic difficulty of the $SD$ models. When the effects of perturbation theory produce large corrections they indicate that the first approximations were not dynamically stable. We mentioned that soft gluon effects 
are not intrinsically present in the non-perturbative quantum mechanical basis we use. Whether or not they are added they do not change the power of $Q^{2}$. We demonstrate this explicitly here
by considering a particular two gluon exchange diagram.

Consider, as an example, the amplitude shown in  Fig. \ref{fig:2gluon}.
The hard scattering contributions of such diagrams have been 
analyzed in \cite{LS1}.
We will extract the $|Q|$ dependence of the amplitude in the endpoint region.
The momenta of the virtual fermions and gluons are
$p_{g_{1}}= k'_{3}-k_{3}$, $ p_{f_{1}}=k'_{2}+k'_{3}-k_{3}$, $p_{g_{2}}=P^{\prime}-P
-k'_{1}+k_{1}$,  $p_{f_{2}}=P-P^{\prime}+k'_{1}$. 
Considering only the endpoint region, it is understood that the momenta 
transferred to the spectator fermions
is soft, hence the gluons and the spectator fermions both have low momenta.
One of the terms in the amplitude is
\begin{equation}
    \label{eq:2g}
    \begin{split}
        A &= \int\limits_{endpoint}[dxdk_{T}][dx'dk'_{T}] (\overline{N}^{\prime}\gamma_{5})_{\gamma^{\prime}}(C^{-1}\slashed{P}^{\prime})
        _{\alpha^{\prime}\beta^{\prime}}\Psi_{123}^{\prime}(k'_{i}) \left[\frac{\gamma^{\mu}(\slashed{p}_{f_{2}}+m)\gamma^{\rho}} 
        {p_{f_{2}}^{2} - m^{2}} \right]_{\gamma^{\prime}\gamma}
        \left[\frac{\gamma^{\lambda}(\slashed{p}_{f_{1}}+m) \gamma_{\rho}}{(p^{2}_{f_{1}}-m^{2})}\right]_{\alpha^{\prime}\alpha}\\
        &\hspace{3cm}\times (\gamma_{\lambda})_{\beta^{\prime}\beta} \times \frac{1}{p^{2}_{g_{1}}}  \times \frac{1}{p^{2}_{g_{2}}}\times 
    (\slashed{P}C)_{\alpha\beta}(\gamma_{5}N)_{\gamma}\Psi_{123}(k_{i}) \\
    &\sim \int\limits_{endpoint}[dxdk_{T}][dx'dk'_{T}] \left[ \overline{N}^{\prime}\gamma_{5}\gamma^{\mu}\frac{\slashed{p}_{f_{2}}+m}
    {p_{f_{2}}^{2} - m^{2}}\gamma^{\rho}\gamma_{5}N\right]
    \left[ \frac{Tr[(C^{-1}\slashed{P}^{\prime})^{T}(\gamma^{\lambda}
    (\slashed{p}_{f_{1}}+m)
    \gamma_{\rho})(\slashed{P}C)\gamma_{\lambda}^{T}]}
{p^{2}_{g_{1}}p^{2}_{g_{2}}(p^{2}_{f_{1}}-m^{2})}\right]\\
& \hspace{10cm}\times\Psi_{123}(k_{i})\Psi_{123}^{\prime}(l_{i})
    \end{split}
\end{equation}
The denominators have the form,
\begin{equation}
    \begin{split}
        (p^{2}_{f_{1}}-m^{2})&\propto (1-x'_{1})x_{3}Q^{2}+
        (\vec{k'}_{T_{1}}+\vec{k}_{T_{3}})^2 + m^{2}\\
        (p^{2}_{f_{2}}-m^{2})& \propto (1-x'_{1})Q^{2}+\vec{k'}_{T_{1}}^{2}+m^{2} 
    \end{split}
\end{equation}
   
The denominator for the soft fermion, which has the momentum $p_{f_{1}}$, has a $(1-x'_{1})x_{3}Q^{2}$
term which is suppressed in the endpoint region. The gluon denominators have similar behavior,
and these terms do not give a $Q^{2}$ dependence in the denominator. The $Q^{2}$ dependence comes from
the hard fermion ($p_{f_{2}}$) denominator which is proportional to $ (1-x'_{1})Q^{2}$.
The dominant Q dependence in the numerator is of the form $Q \times Q \times (1-x'_{1})Q \times Q$. 
Evaluating this term in the endpoint region using our wave function we obtain
\begin{equation}
    A \sim \int\limits_{endpoint}[dxdk_{T}][dx'dk'_{T}][\overline{N}^{\prime}\gamma^{\mu}N]\frac{(1-x'_{1})Q^{4}}
    {(1-x'_{1})Q^{2}}x'_{1}(1-x'_{1})x_{1}(1-x_{1}) \propto [\overline{N}^{\prime}\gamma^{\mu}N]\times
    \frac{1}{Q^{4}}
\end{equation}
Hence we obtain the expected momentum dependence.

\begin{spacing}{1}
\begin{small}

\end{small}
\end{spacing}

\end{document}